\documentclass[12pt]{article}

\usepackage{graphics,epsfig}
\usepackage{amsbsy}
\usepackage{amsfonts}
\usepackage{amsmath}
\usepackage{graphicx}

\def\XXint#1#2#3{{\setbox0=\hbox{$#1{#2#3}{\int}$ }
\vcenter{\hbox{$#2#3$ }}\kern-.6\wd0}}

\def\id{\protect{{1 \kern-.28em {\rm l}}}}

\newcommand{\beq}{\begin{equation}}
\newcommand{\eeq}{\end{equation}}
\newcommand{\beqr}{\begin{displaymath}}
\newcommand{\eeqr}{\end{displaymath}}
\newcommand{\beqa}{\begin{eqnarray}}
\newcommand{\eeqa}{\end{eqnarray}}
\newcommand{\beqar}{\begin{eqnarray*}}
\newcommand{\eeqar}{\end{eqnarray*}}

\def\k{\kappa}

\def\p{{\partial}}
\def\nn{\nonumber}

\newcommand{\adss}[2]{\mbox{$AdS_{#1}\times {S}^{#2}$}}

\newcommand{\bino}[2]{\ensuremath{\left(\begin{array}{c} \scriptstyle #1 \\ \scriptstyle #2 \end{array}\right)}}

\def\dalemb#1#2{{\vbox{\hrule height .#2pt
        \hbox{\vrule width.#2pt height#1pt \kern#1pt
                \vrule width.#2pt}
        \hrule height.#2pt}}}

\def\half{\frac{1}{2}}
\let\a=\alpha \let\b=\beta \let\g=\gamma \let\d=\delta \let\e=\epsilon
\let\z=\zeta  \let\th=\theta  \let\k=\kappa
\let\l=\lambda \let\m=\mu  \let\x=\xi \let\p=\pi 
\let\s=\sigma \let\t=\tau   \let\c=\chi 
 \let\vep=\varepsilon
\let\w=\omega      \let\G=\Gamma \let\D=\Delta \let\Th=\Theta \let\L=\Lambda
 \let\P=\Pi \let\S=\Sigma  
\let\C=\Chi \let\W=\Omega
\let\la=\label \let\ci=\cite 
  
\def\nn{\nonumber} \def\bd{\begin{document}} \def\ed{\end{document}}
\def\ds{\documentstyle} \let\fr=\frac \let\bl=\bigl \let\br=\bigr
\let\Br=\Bigr \let\Bl=\Bigl
\let\bm=\bibitem
\let\na=\nabla
\def\tU{{\widetilde U}}
\let\pa=\partial \let\ov=\overline
\def\ie{{\it i.e.\ }}
\newcommand{\be}{\begin{equation}}
\newcommand{\ee}{\end{equation}}
\def\ba{\begin{array}}
\def\ea{\end{array}}
\def\ft#1#2{{\textstyle{{\scriptstyle #1}\over {\scriptstyle #2}}}}
\def\fft#1#2{{#1 \over #2}}
\def\F#1#2{{ F_{#1}^{(#2)} }}
\def\cF#1#2{{ {\cal F}_{#1}^{(#2)} }}

\def\={\, =\, }
\def\+{\, +\, }
\def\-{\, -\, }

\def\R{{\bf R}}
\def\sst#1{{\scriptscriptstyle #1}}
\def\oneone{\rlap 1\mkern4mu{\rm l}}
\def\e7{E_{7(+7)}}
\def\td{\tilde}
\def\wtd{\widetilde}
\def\im{{\rm i}}
\newcommand{\ho}[1]{$\, ^{#1}$}
\newcommand{\hoch}[1]{$\, ^{#1}$}
\newcommand{\bea}{\begin{eqnarray}}
\newcommand{\eea}{\end{eqnarray}}
\newcommand{\ra}{\rightarrow}
\newcommand{\lra}{\longrightarrow}
\newcommand{\Lra}{\Leftrightarrow}
\newcommand{\ap}{\alpha^\prime}
\newcommand{\bp}{\tilde \beta^\prime}
\newcommand{\cB}{{\cal B}}
\newcommand{\cO}{{\cal O}}
\newcommand{\vecx}{\vec{x}}
\newcommand{\vecy}{\vec{y}}
\newcommand{\vecp}{\vec{p}}
\newcommand{\vecq}{\vec{q}}
\newcommand{\tr}{{\rm tr} }
\newcommand{\Tr}{{\rm Tr} }
\newcommand{\eg}{{\it e.g.}}

\newcommand{\cL}{{\cal L}}
\newcommand{\cA}{{\cal A}}
\newcommand{\cD}{{\cal D}}
\def\sst#1{{\scriptscriptstyle #1}}

\def\ve{\varepsilon}
\def\vf{\varphi}
\def\F{\Phi}
\def\wg{\wedge}

\def \foot {\footnote}
\def \bi{\bibitem}

\def \tr {{\rm tr}}
\def \ha {{1 \over 2}}
\def \td {\tilde}
\def \ci{\cite}
\def \N {{\mathcal N}}
\def \ww {\Omega}
\def \const {{\rm const}}
\def \ss {\sum_{i=1}^3 }
\def \t {\tau}
\def\S{{\mathcal S} }
\def \nn {\nu}
\def \XX {{\rm X}}

\def \lra {\leftrightarrow}
\def \vom {{\bar \omega}}
\def \E {{\mathcal  E}} \def \J {{\mathcal  J}}
\def \YY {{\rm Y}}

\def \d {\del}
\def \rJ {{J}}
\def \sms {sigma models\ }
\def \sm {sigma model\ }
\def \L {\Lambda}
\def \gl {\ell}
\def \tr {{\rm tr\ }}
\def\z{\zeta}
\def\zi{\zeta_1}
\def\zii{\zeta_2}
\def\K{\mbox{K}}
\def\eE{\mbox{E}}   \def \vt {\vartheta}
\def \vr {\varrho}
\def \wup {w}

\def\dg{\dagger}
\def\a{\alpha}
\def\b{\beta}
\def\e{\varepsilon}
\def\p{\phi}
\def\ap{\alpha^\prime}
\def\I{{\cal I}}

\def\R{{\bf R}}
\def\Z{{\bf Z}}
\def\C{{\bf C}}
\def\P{{\bf P}}
\def\xb{{\bar X}}
\def\Tr{{\rm  Tr}}
\def\tr{{\rm  tr}}

\def \del{\partial}
\def \a {\alpha}
\def \aa {{\a'}}
\def\g{\gamma}
\def\s{\sigma}
\def\z{\zeta}
\def\zi{\zeta_1}
\def\zii{\zeta_2}
\def\ov{\over}

\def\I{{\cal I}}
\def\J{{\mathcal J}}
\def \ok {{1\ov \k}}
\def\LL{{\mathcal L }}
\def \jL {{J}}
\def \om {\omega}
\def \cL {{\mathcal L}} \def \cH {{\mathcal H}}
\def\E{{\mathcal E}}
\def\w{\omega}
\def\b{\beta}
\def\l{\lambda}
\def\eps{\epsilon}
\def\vep{\varepsilon}
\def \De {{\mathcal D}}

 \def \cV {{\cal V}}
\def  \Jt {  {J}_{\rm tot}    }

\def \k {\kappa}
\def\foot{\footnote}
\def \four{{\textstyle {1\ov 4}}}
 \def \third { \textstyle {1\ov 3
}}
\def\det{\hbox{det}}
\def \ci {\cite}

\def \foot {\footnote}
\def \bi{\bibitem}

\def \tr {{\rm tr}}
\def \ha {{1 \over 2}}
\def \tid {\tilde}
\def \vv {{\rm v}}
\def \tl {{\tilde \l}}
\def \XX {{\rm X}}
\def \ta {{\tilde \a}}
\def \fo { {1\ov 4}}
\def \ep {\epsilon}
\def \inti {{\int^{2\pi}_0 {d \sigma \ov 2 \pi}}}

\def \d {\partial}
\def \K {{\rm S}}
\def \el {\ell}
\def \Tr {{\rm Tr}}
\def \P {\Phi}
\def \l  {\lambda}
\def \tl {{\tilde \l}}
\def \bl {{\tilde \l}}
\def \const {{\rm const}}
\def \V {v}

\def \bv {v^*}
\def \vv {{\rm v}}
\def \LL {{\mathcal L}}
\newcommand{\PV}[1]{P_{\!\!_{V_{#1}}}}

\def \bL {\ell}
\def \M {{\mathcal M}}
\def \N {{\mathcal N}}
\def \S {{\rm S}}
\def \vn {\vec n}
\def \tl {\td \l}
\def \td {\tilde}
\def \Prod {\Pi}
\def \O {{\mathcal O}}
\def \Q {{\rm  Q}}
\def \D {\Delta}
\def \N {{\mathcal N}}
\def\tN{{\tilde N}}

\def \m {\mu}
\def \vs {\vec \s}
\def \ie {i.e.}

\def \cD {{\cal D}}

\def \rS {{\rm S}}
\def\as{{\a}}
\newcommand{\bra}[1]{\mbox{$\langle #1 |$}}
\newcommand{\ket}[1]{\mbox{$| #1 \rangle$}}

\newcommand{\Gg}{G}

\newcommand{\auth}{AUTHORS}

\def\thb{\bar{\theta}}
\def\Thb{\bar{\Theta}}
\def\barp{\bar{p}}
\def\barq{\bar{q}}
\def\barc{\bar{c}}
\def\bard{\bar{d}}
\def\e{\epsilon}

\def \bi{\bibitem}
\def \la {\label}

\def \l {\lambda}
\def\foot{\footnote}
\def \tl  {{\tilde \l}}
\def \sql {{\sqrt \l}}
\def \adss {$AdS_5 \times S^5$\ }
\newcommand{\rf}[1]{(\ref{#1})}
\def \ov {\over}

\def\th{\theta}
\def\Th{\Theta}
\def\vth{\vartheta}

\def\vth{\vartheta}

\def\ra{\rightarrow}
\def\N{{\cal N}}
\def\F{{\cal F}}
\def\cc{\circ}
\def\eqv{\equiv}

\def\ni{\noindent}
\def \ha{{1\ov 2}}
\def \bw {{\rm w}}

\def\r{{\rm r}}

\def \cT {{\cal T}}
\def \no {\nonumber}

\def \J {\mathcal{J}}
\def \del {\partial}

\def \bps {{\bar \psi}}
\def \sqbl {\sqrt{\bar \lambda}}

\def\dF{\dot{F}}
\def\dG{\dot{G}}
\def\df{\dot{f}}
\def \E {{\cal E}}

\def \S {{\cal S}}
\def \J {{\cal J}}

\def\ms{\mathcal{S}}
\def\mj{\mathcal{J}}
\def\soj{\fr{\ms}{\mj}}
\def \R {{\bf R}}
\def \om {\omega}
\def \tH {\widetilde H}
\def \bE {\bar E}
\def \x {{\cal X}}

\def \hV {{\hat V}}

 \def \bb {\bar \beta}
\def \W {{\cal E}}

\def \bi{\bibitem}
\def \la {\label}

\def \l {\lambda}
\def\foot{\footnote}
\def \tl  {{\tilde \l}}
\def \sql {{\sqrt \l}}
\def \sqtl {{\sqrt {\tilde \l}}}

\def \HH {{\rm E}}
\def \cS {{\cal S}}
\def \cL {{\cal L}}

\def \adss {$AdS_5 \times S^5$\ }

\def \D {\Delta}
\def \thet {\theta}
 \def \t {\tau}
 \def \p {\phi}
 \def \r {\rho}
 \def \rN {{\rm N}}
 \def\tw{{\tilde w}}
 \def\hJ{{J}}
 \def\hw{{w}}
 \def\hl{{\lambda}}
 \def\hth{{\theta}}
 \def\NN{{\cal N}}
 \def \bv {{ \bar w}}
\def \vn {{\vec n}}
\newcommand{\sfrac}[2]{{\textstyle\frac{#1}{#2}}}
\def \bl {{ \bar \lambda}}
\def \bp {{\bar p}}
\def \bu {{\bar u}}
\def \sha {\sfrac{1}{2}}
\def \w {\omega}
\def \ov {\over}
\def \vl { \vec \ell}
\def \varpi {{\rm w}}
\def \OO {{\cal O}}
\def \bG {\bar \G}

\def \c {\gamma}

\def \ss {{\rm s}}

\def \ve {\varepsilon}
\def \pa{\partial}
\def \I {{\cal I}}
\def \LL {{\cal L}}
\def \ep {\epsilon}
\def \R {{\rm R}}
\def \tilt {{\tilde t}}

\def\pic #1#2{\hbox{\lower#1pt\hbox{~\mbox{\epsfxsize=20truemm \epsffile{#2}}}}}

\def\pic #1#2#3{\hbox{\lower#1pt\hbox{~\mbox{\includegraphics[scale=#3]{#2}}}}}

\def \bt {\bar\theta}
\def \te {\theta}

\def \cc {{\rm f}}
\def \d {\delta}

\def \cL {{\cal L}}
\def \S  {{\cal S}}
\def \pp {{q}}
\def \vt {\vartheta}
\def \mm {{\cal  \ell}}
\def \Z {{\cal Z}}
\def \pa {\partial}
\def \C {{\cal C}}
\def \be {\bea}
\def \ee {\eea}
\def \c {\gamma}  \def \d {\delta}
\def \eps {\epsilon}

\def \bp {\begin{pmatrix}}  \def \ep {\end{pmatrix}}

 \def \T {{\cal T}}

\def \bp {\begin{pmatrix}}  \def \epm {\end{pmatrix}}
\def \ha {{\textstyle{1 \ov 2}}}

\begin{document}
\overfullrule=0pt
\parskip=2pt
\parindent=12pt
\headheight=0in \headsep=0in \topmargin=0in \oddsidemargin=0in

\vspace{ -3cm} \thispagestyle{empty} \vspace{-1cm}
\begin{center}
 \vspace{2cm}
{\Large\bf
 On the dressing phase in the SL(2) Bethe Ansatz
\\\vspace{0.3cm}
 }

\vspace{.5cm} {
  M. Kruczenski \footnote{markru@purdue.edu} and A. Tirziu \footnote{atirziu@purdue.edu}\\
\vskip 0.3cm

{\em
Department of Physics, Purdue  University,\\
W. Lafayette, IN 47907-2036, USA.}}

\end{center}

 \begin{abstract}
In this paper we study the function $\chi(x_1,x_2,g)$ that determines the dressing phase that appears in the all-loop Bethe
Ansatz equations for the $SL(2)$ sector of $\mathcal{N}=4$ super Yang-Mills theory. First, we consider the coefficients $c_{r,s}(g)$
of the expansion of $\chi(x_1,x_2,g)$ in inverse powers of $x_{1,2}$. We obtain an expression in terms of a single integral valid for all
values of the coupling $g$. The expression is such that the small and large coupling expansion can be simply computed in agreement with
the expected results. This proves the, up to now conjectured, equivalence of both expansions of the phase. The strong coupling expansion
is only asymptotic but we find an exact expression for the value of the residue which can be seen to decrease exponentially with $g$.

After that, we consider the function $\chi(x_1,x_2,g)$ itself and, using the same method, expand it for small and large coupling.
All small and large coupling coefficients $\chi^{(n)}(x_1,x_2)$, for even and odd $n$, are explicitly given in terms of finite sums
or, alternatively, in terms of the residues of generating functions at certain poles.

\end{abstract}

\newpage


\section{Introduction}

Much of the recent progress in understanding the connection between the strong and weak coupling   regimes
in the $\N=4$ SYM was achieved using the Bethe ansatz technique within the AdS/CFT correspondence \cite{malda}.
Not only is this useful in the gauge theory, but it should also give the spectrum
of the free strings in the $AdS_5 \times S^5$ background.

While at weak coupling the Bethe ansatz description appeared as appropriate for the associated spin chain \cite{MZ}, at strong coupling, namely in the dual string theory picture, it was first proposed at the classical level
in \cite{afs}. An important ingredient in the all loop Bethe ansatz equations is the coupling dependent
dressing phase. Using $1$-loop string results near particular states \cite{ptt}, a few leading terms
 in the 1-loop dressing phase were found in \cite{bt}. In \cite{hl, FK} the full 1-loop strong coupling expansion of the coupling dependent coefficient $c_{r,s}(g)$ entering the dressing phase was found and tested. Further progress in understanding the next orders in strong coupling expansion of $c_{r,s}(g)$ was made in \cite{j} which lead to the finding of all strong coupling expansion coefficients of $c_{r,s}(g)$ \cite{bhl}.

  For the particular $SL(2)$ sector, the all loop Bethe ansatz (BES)
equations were proposed in \cite{bes}.  This ansatz is only asymptotic as it is only supposed to work for large values of the length of the spin chain $J$.
The all loop Bethe ansatz was tested by computing the one-cut large $S$ anomalous dimensions for the operators of the type $\tr(\Phi D_{+}^S \Phi)$.   More specifically, the all loop anomalous dimension was shown to be \cite{K, km, es,bes}
\begin{equation}
E- S= f(\lambda) \ln S + \mathcal{O}(S^{0})  \label{yat}
\end{equation}
The all loop Bethe ansatz equations for this solution lead to an integral equation for the universal function $f(\lambda)$. The  weak coupling expansion of $f(\lambda)$ was checked at weak coupling to four loops against a direct gauge theory computation \cite{bcdks}. The function $f(\lambda)$ is related to the cusp anomaly of light-like Wilson loops \cite{K, km} which can also be computed at strong coupling using AdS/CFT \cite{kru, m}. The logarithmic scaling was studied at weak
and strong coupling \cite{BGK, FTT, CK}.
The complicated integral equation for $f(\lambda)$ obtained in \cite{bes} was solved at strong coupling in \cite{bkk}. Remarkably, it matches the expansion obtained directly on the string side to two loops in strong coupling expansion \cite{ft1,rtt}. The validity of the asymptotic all loop Bethe ansatz was checked to next order in large $S$ expansion \cite{fz, fgr} for the folded string solution corresponding to twist two operators, and also for a class of more complex solutions, namely  the spiky string solutions \cite{fkt}.

Although the asymptotic Bethe ansatz was tested for certain states in the $SL(2)$ sector, a direct rigorous proof of the conjectured relationship between the strong and weak coupling expansions of $c_{r,s}(g)$ for all $r,s$ was not obtained. For $r=2,s=3$ a proof was obtained in \cite{bes}. A further attempt to prove the relationship was made in \cite{gh} only for certain coefficients of the expansion of $c_{2,s}$. In \cite{kl, fr} a relationship between the weak and strong coupling expansions of $c_{r,s}(g)$ was found but the strong coupling expansion was treated in a non rigorous way. It is the goal of this paper to study in detail the properties of $c_{r,s}(g)$, and to give a proof of the conjectured expansions for all $r,s$.

Starting with the weak coupling expansion of $c_{r,s}(g)$ we sum the series, and then we obtain a single integral representation formula in the complex plane for $c_{r,s}(g)$. This allows us to systematically analyze weak and strong coupling expansions by simply deforming the contour as appropriate for the expansion we want. As was pointed out already in \cite{bes}, we show that at strong coupling the expansion of $c_{r,s}(g)$ is an asymptotic  series. We obtain a well defined integral for the remainder, which can be evaluated in principle as precise as desired.  We estimate the remainder integral to behave exponentially as $g^{-3/2} e^{-8 \pi g}$. Exponential behavior was obtained for the cusp anomaly $f(\lambda)$ at strong coupling in \cite{bk}. The non-perturbative scale obtained in \cite{bk} is consistent with the mass gap of the two-dimensional bosonic O(6) sigma model embedded into the $AdS_5 \times S^5$ string theory \cite{am1}.

As a byproduct of the integral representations that we find, we are able to sum the dressing phase and therefore obtain all weak and strong coupling expansions of the dressing phase in terms of finite sums, which can be readily be performed at any order as needed. In addition, we found explicitly the exponentially suppressed part of the dressing phase. It would be interesting to use these results in the computation of the function $f(\lambda)$, especially to recover the non-perturbative scale obtained in \cite{bk}.

The paper is organized as follows. In section 2 we review the all loop dressing phase  in the $SL(2)$ Bethe ansatz, as well as the expansions of $c_{r,s}(g)$ proposed in \cite{bes}. In section 3 we find a double integral representation of $c_{r,s}$ by summing the weak coupling expansion. The main result of this paper, i.e. a single integral representation of $c_{r,s}$ suitable for any expansion is obtained in section 4. In section 5 we study the properties of $c_{r,s}$, perform weak/strong coupling expansions, and thus prove the relationship  between them. The summing of the dressing phase and its $g$ expansions are done in section 6. Finally, in section 7 we present a summary of the results while in Appendix A we check by a different method the expansions of a simple illustrative example that we use.

\section{Dressing phase}

As mentioned, the phase $\theta(x^\pm_1,x^\pm_2)$ defined as
\begin{equation}
\theta(x^\pm_1,x^\pm_2)=\sum_{r=2}^{\infty}\sum_{s=r+1}^{\infty} c_{r,s}(g) [q_r (x_1^{\pm}) q_s (x_2^{\pm})-q_s (x_1^{\pm}) q_r (x_2^{\pm})]
\end{equation}
plays an important role in the all-loop Bethe-ansatz. Here
\begin{equation}
q_r = \frac{i}{r-1}\bigg(\frac{1}{(x^{+})^{r-1}}-\frac{1}{(x^{-})^{r-1}}\bigg)
\end{equation}
and the coefficients $c_{r,s}(g)$ are given below. $x^\pm(u)$ are $g$-dependent and defined through $u\pm\tfrac{i}{2}=x^\pm(u)+\tfrac{g^2}{x^\pm(u)}$.

It is convenient to write the dressing phase in the following way \cite{bhl}
\begin{eqnarray}
\theta(x^\pm_1,x^\pm_2,g) &=& \chi(x_1^{+},x_2^{+},g)-\chi(x_1^{+},x_2^{-},g)-\chi(x_1^{-},x_2^{+},g)+\chi(x_1^{-},x_2^{-},g)\nonumber\\
&-&\chi(x_2^{+},x_1^{+},g)+\chi(x_2^{-},x_1^{+},g)+\chi(x_2^{+},x_1^{-},g)-\chi(x_2^{-},x_1^{-},g)
\end{eqnarray}
where
\begin{equation}
\chi(x_1,x_2,g)=-2 \sum_{r=2}^{\infty}\sum_{s=r+1}^{\infty}  \frac{\tilde{c}_{r,s}(g)}{x_1^{r-1} x_2^{s-1}}
\end{equation}
and, for convenience, we defined
\beq
\tilde{c}_{r,s}(g) = \half \frac{1}{(r-1)(s-1)} c_{r,s}(g),
\eeq
  It turns out that the coefficients $c_{r,s}$ vanish unless $r+s$ is odd. Therefore we can define two integers
\beq
 m = \half(r+s-3), \ \ \ \bar{m} = \half(s-r-1), \ \ \ \ s=m+\bar{m}+2, \ \ \ \  r=m-\bar{m}+1, \quad m \geq \bar{m}+1
\label{mmbar}
\eeq
and express $\chi(x_1,x_2,g)$ as
\beq
\chi(x_1,x_2,g) =-2 \sum_{\bar{m}=0}^{\infty}\sum_{m=\bar{m}+1}^{\infty} \frac{\tilde{c}_{m,\bar{m}}(g) }{x_1^{m-\bar{m}} x_2^{m+\bar{m}+1}}
\eeq
where
\beq
\tilde{c}_{m,\bar{m}}(g) =  \tilde{c}_{r,s}(g), \ \ \ \ r=m-\bar{m}+1, \quad m \geq \bar{m}+1
\eeq
By a slight abuse of notation we still call the coefficients as $\tilde{c}$. To avoid confusion, from now on we are going to use always the notation
$\tilde{c}_{m,\bar{m}}$, or equivalently
\beq
c_{m,\bar{m}} = 2(m-\bar{m})(m+\bar{m}+1) \tilde{c}_{m,\bar{m}}
\label{cct}
\eeq
For small coupling $g<\frac{1}{4}$ the coefficients $\tilde{c}_{m,\bar{m}}(g)$ can be expanded\footnote{We define the coefficients as the straight-forward expansion of $\chi(x_1,x_2,g)$. As a result there is an overall minus sign with respect to \cite{bes}} in powers
of $g$:
\beq
\tilde{c}_{m,\bar{m}}(g) = \sum_{k=1}^\infty \tilde{c}^{(2k)}_{m,\bar{m}}\, g^{2k+1}
\eeq
 From \cite{bes}, we find that the coefficients  $\tilde{c}^{(n)}_{m,\bar{m}}$, have a nice symmetric form:
\beq
 \tilde{c}^{(2k)}_{m,\bar{m}}  = \frac{(-)^{k+m+\bar{m}} \zeta(1+2k)}{(2+2k)^2(1+2k)B(1-m+k,2+m+k)B(1-\bar{m}+k,2+\bar{m}+k)}
\label{CS}
\eeq
where $B(x,y)=\frac{\Gamma(x)\Gamma(y)}{\Gamma(x+y)}$ is Euler's beta function.
 These coefficients also have an asymptotic expansion for large $g$ as
\beq
 \tilde{c}_{m,\bar{m}}(g) = \sum_{n=0}^N \tilde{c}^{(-n)}_{m,\bar{m}}\, g^{1-n} +R_N
\label{lec}
\eeq
 where, for $n>1$
\beq
 \tilde{c}^{(-n)}_{m,\bar{m}} = \frac{\zeta(n)}{2(-2\pi)^n \Gamma(n-1)} \
      \frac{\Gamma\left(m+\half n\right)\Gamma\left(\bar{m}+\half n\right)}{\Gamma\left(m+2-\half n\right)\Gamma\left(\bar{m}+2-\half n\right)}
\label{CL}
\eeq
and
\beq
\tilde{c}^{(0)}_{m,\bar{m}} = \frac{1}{2m(m+1)} \delta_{\bar{m},0}, \ \ \ \tilde{c}^{(-1)}_{m,\bar{m}} =-\frac{1}{\pi}\frac{1}{(2m+1)(2\bar{m}+1)}
\label{CL01}
\eeq
Since the expansion is only asymptotic we should sum a finite number of terms and include a residue $R_N$ to have an equality between both sides of eq. (\ref{lec}).
 Obviously it is quite important that both, small and large coupling expansions correspond to the same function. However, up to know, this was only a conjecture except for
$\bar{m}=0$, $m=1$ (or $r=2$, $s=3$) where it was proved in \cite{bes}. Moreover, since the strong coupling expansion is only asymptotic, it
is important to give an expression for the residue $R_N$ so that we can estimate the error. In the following we give a proof of the equivalence of
both expansions by considering an expression valid for all values of the coupling and such that it can be easily expanded at large and small $g$ with the expected results.
It also provides an exact expression for the residue $R_N$.

Moreover, we extend these results to the function $\chi(x_1,x_2,g)$. Such function can also be expanded in powers of $g$ as
\beqa
\chi(x_1,x_2,g) &=& \sum_{k=1}^\infty \chi^{(2k)}(x_1,x_2) g^{2k+1}, \ \ \ \ g<\frac{1}{4}  ,\\
\chi(x_1,x_2,g) &=& \sum_{n=0}^N \chi^{(-n)}(x_1,x_2) g^{1-n} + R_N, \ \ \ \ g\rightarrow\infty  .
\eeqa
Again, the second expansion is only asymptotic. For given $g$ there is an optimal value of $N$ such that the residue $R_N$ is smallest. In this paper we find explicit expressions for
all the coefficients of such expansion as well as for the residue $R_N$. The coefficients are in terms of finite sums or alternatively in terms of the residue of given functions at certain poles.

\section{A double integral representation for $c_{m,\bar{m}}(g)$}

 As a first step we are going to construct a generating function for the coefficients  $\tilde{c}^{(2k)}_{m,\bar{m}}$ of the small coupling expansion.
When computing $\chi(x_1,x_2,g)$ we only need to consider $m>\bar{m}\ge0$ but there is nothing wrong with extending the formulas to all values of $m,\bar{m}$.
In fact, for fixed $n=2k$, if we vary $m$ (or $\bar{m}$) most coefficients vanish, the only ones that survive are such that
\beq
 -k-1 \leq m \leq k, \ \ \ \ -k-1 \leq \bar{m} \leq k .
\eeq
 We can therefore define a double periodic generating function
\beq
\tilde{C}^{(n)}(\mu,\nu) = \sum_{m,\bar{m}=-\infty}^{\infty} \tilde{c}^{(n)}_{m,\bar{m}} e^{2im\mu+2i\bar{m}\nu} .
\eeq
Notice that the series trivially converges since it actually has a finite number of terms.
Given $\tilde{C}^{(n)}(\mu,\nu)$ we can recover the coefficients by Fourier analysis. Now we need to compute
\beq
\sum_{m=-k-1}^{k} (-)^m \frac{e^{2im\mu}}{B(1-m+k,2+m+k)} = 2^{2k+1}(2k+2) i e^{-i\mu} (\sin\mu)^{1+2k}
\eeq
We then get
\beq
\tilde{C}^{(n)}(\mu,\nu) = (-)^{k+1} \frac{\zeta(1+2k)}{1+2k} 4^{2k+1} e^{-i\mu-i\nu} (\sin\mu\sin\nu)^{2k+1}
\eeq
 Now we can sum over $k$ and define
\beq
 \tilde{C}(\mu,\nu;g) =  \sum_{k=1}^{\infty} \tilde{C}^{(2k)}(\mu,\nu) g^{2k+1}
                      =  - e^{-i\mu-i\nu} \sum_{k=1}^{\infty} (-)^k  \frac{\zeta(1+2k)}{1+2k} (4g\sin\mu\sin\nu)^{2k+1}
\eeq
 This sum can be done explicitly and we get
\beqa
 \tilde{C}(\mu,\nu;g) &=&  -e^{-i\mu-i\nu} \left[ -\gamma \bar{g} + \frac{i}{2} \ln\left(\frac{\Gamma(1+i\bar{g})}{\Gamma(1-i\bar{g})}\right)\right] \\
              &=& e^{-i\mu-i\nu} \left[ \gamma \bar{g} + \arg(\Gamma(1+i\bar{g})) \right]
\eeqa
where we defined
\beq
\bar{g} = 4 g \sin\mu\sin\nu
\eeq
The statement is that if one expands this last function in powers of $\bar{g}$ and then Fourier analyze it in $\mu,\nu$, the coefficients, by construction,
are precisely the $c^{(n)}_{m,\bar{m}}$ at small coupling. The function $\tilde{c}_{m,\bar{m}}(g)$ can be obtained as
\begin{equation}
\tilde{c}_{m,\bar{m}}(g)=\int_{0}^{\pi}\frac{d \mu}{\pi} \int_{0}^{\pi}\frac{d \nu}{\pi}e^{-2 i \mu m - 2 i \nu \bar{m}} \tilde{C}(\mu,\nu; g)
\label{cmmbar}
\end{equation}
If we wish, from eq.(\ref{cct}), we can also find the coefficients $c_{m,\bar{m}}$ as
\beq
C(\mu,\nu;g) =
 2 \left(-\frac{1}{4}\partial_\mu^2 -\frac{i}{2}\partial_\mu  + \frac{1}{4}\partial_\nu^2 +\frac{i}{2}\partial_\nu \right) \tilde{C}(\mu,\nu,g)
\eeq
Some algebra gives
\beq
 C(\mu,\nu;g) = -8 g^2 (\sin^2\nu-\sin^2\mu)\ \partial_{\bar{g}}^2 \tilde{C}(\mu,\nu,g)
\eeq
We finally get
\beq
 C(\mu,\nu;g) = 8 g^2 e^{-i\mu-i\nu} (\sin^2\nu-\sin^2\mu)\ \mbox{Im} \psi'(1+i\bar{g})
\label{genfun}
\eeq
where $\psi'$ denotes the derivative of the $\psi$ function, $\psi(x)=\Gamma'(x)/\Gamma(x)$.
Therefore
\begin{equation}
c_{m,\bar{m}}(g)=\int_{0}^{\pi}\frac{d \mu}{\pi} \int_{0}^{\pi}\frac{d \nu}{\pi}e^{-2 i \mu m - 2 i \nu \bar{m}} C(\mu,\nu; g)
\label{gen2}
\end{equation}
Of course given $\tilde{c}_{m,\bar{m}}$ we can get $\tilde{c}_{m,\bar{m}}$ multiplying by the corresponding factor (\ref{cct}) and vice-versa,
the purpose of deriving the last equation is that it is somewhat easier to work with the generating function $C(\mu,\nu;g)$.
We are interested now in finding the strong coupling expansion. If we naively try to expand $\psi'(1+i\bar{g})$ in eq.(\ref{genfun})
for large $\bar{g}$ we find that the resulting integrals over $\mu$ and $\nu$ diverge. The reason being that we need $\bar{g}=4g\sin\mu\sin\nu$
to be large but, even if $g$ is large, close to $\mu=0$ or $\nu=0$ we can have $\bar{g}$ as small as we want.  Notice that in \cite{bes}, the
alternative expression in term of Bessel functions (here we use our sign convention and redefine the indices according to eq.(\ref{mmbar}))
\beq
\tilde{c}_{m,\bar{m}} = \cos(\pi \bar{m}) \int_0^\infty dt \frac{J_{m-\bar{m}}(2gt)J_{m+\bar{m}+1}(2gt)}{t(e^t-1)} \label{bessel}
\eeq
was given. However such expression does not give an obvious large $g$ expansion either since we cannot assume that $2gt$ is large around $t=0$.

In the next section we derive an alternative expression for the coefficients valid for all $g$ and which allows for a simple expansion,
both at large and small $g$.

\section{A single integral representation formula for $c_{m, \bar{m}}(g)$}

Consider an integral of the type
\beq
f(g)= \int_0^{\pi} d\mu  \int_0^{\pi} d\nu\ \sin\mu\sin\nu F(\mu,\nu)\, \mbox{Im}\psi'(1+4ig\sin\mu\sin\nu)
\label{fdef}
\eeq
as we had in the previous section.  Using the following integral representation
\begin{equation}
\psi'(x)=\int_0^{\infty}dt \frac{t e^{-x t}}{1-e^{-t}}
\end{equation}
we can rewrite $f(g)$ as:
\beq
f(g) = \int_0^\infty d\zeta \sin(4g\zeta) H(\zeta)
\eeq
with
\beq
H(\zeta) = -\zeta \int_0^{\pi} \int_0^{\pi} \frac{d\mu d\nu}{\sin\mu\sin\nu} \frac{F(\mu,\nu)}{e^{\frac{\zeta}{\sin\mu\sin\nu}}-1}
\eeq
Notice the integral converges because near $\mu=0,\pi$ or $\nu=0,\pi$ there is an exponential suppression ($\zeta>0$).
Now let us manipulate this integral
\beq
H(\zeta) = - \zeta \int_0^\infty du \int_0^{\pi} \int_0^{\pi} \frac{d\mu d\nu}{\sin\mu\sin\nu}
          \delta(u-\sin\mu\sin\nu) \frac{F(\mu,\nu)}{e^{\frac{\zeta}{\sin\mu\sin\nu}}-1}
\eeq
Which is the same since the integral over $u$ is 1. Since $u>0$ we can use
\beq
\delta(u-u_0)=\frac{1}{u}\delta(\ln u-\ln u_0) = \frac{1}{u} \int_{-\infty}^{+\infty} \frac{d\eta}{2 \pi}\, e^{-i\eta(\ln u -\ln u_0)}
             = \frac{1}{u} \int_{-\infty}^{+\infty} \frac{d\eta}{2 \pi} \, \left(\frac{u_0}{u}\right)^{i\eta+c}
\eeq
where $c$ is an arbitrary real number $0< c <1$ that we introduce to ensure the convergence of the integral below. It is arbitrary since the
delta function assures $u=u_0$ and therefore $(u/u_0)^c=1$.
We get
\beq
H(\zeta) = -\zeta \int_{-\infty}^{+\infty} \frac{d\eta}{2 \pi} \int_0^\infty du \int_0^{\pi} d\mu \int_0^{\pi} d\nu
 \frac{u^{-i\eta-2-c}}{e^{\frac{\zeta}{u}}-1} \left(\sin\mu\sin\nu\right)^{i\eta+c} F(\mu,\nu)
\eeq
Now define
\beq
\phi(s) = \int_0^{\pi} d\mu \int_0^{\pi} d\nu  \left(\sin\mu\sin\nu\right)^{s} F(\mu,\nu)
\label{phidef}
\eeq
and compute
\beq
\int_0^\infty du \frac{u^{-i\eta-2-c}}{e^{\frac{\zeta}{u}}-1}
= \zeta^{-i\eta-1-c}\int_0^\infty dx\frac{x^{i\eta+c}}{e^x-1} = \zeta^{-i\eta-1-c} \Gamma(1+c+i\eta)\zeta(1+c+i\eta)
\eeq
which converges in virtue of $c>0$. We therefore obtain
\beq
H(\zeta) = - \int_{-\infty}^{+\infty} \frac{d\eta}{2\pi}  \zeta^{-i\eta-c} \Gamma(1+c+i\eta)\zeta(1+c+i\eta)\phi(i\eta+c)
\eeq
We still need now to do the (sine) Fourier transform
\beq
\int_0^\infty d\zeta \sin (4 g \zeta) \zeta^{-i\eta-c} = (4g)^{i \eta +c-1} \cos \frac{\pi (i \eta+c)}{2}\Gamma(1-c-i \eta)
\label{sineF}
\eeq
Notice that this integral is well defined for $g>0$ and $0<c<1$ \footnote{In fact it is well defined for $0<c<2$ however since $c$ is arbitrary we take
it between zero and one. This simplifies the notation further on and also keeps the formula valid if we use the real part of $\psi'$ instead of the imaginary
part in eq.(\ref{fdef}). We just need to replace sine by cosine in (\ref{sineF}).}.
Finally we obtain
\beqa
f(g) &=& - \int_{-\infty}^\infty \frac{d\eta}{2 \pi} (4g)^{ i \eta+c-1}
        \cos \frac{\pi (i \eta+c)}{2} \zeta(1+i\eta+c)\Gamma(1-i\eta-c)\Gamma(1+i\eta+c)\phi(i\eta+c) \\
     &=& i \int_{c-i\infty}^{c+i\infty} \frac{ds}{2 \pi}\, (4g)^{s-1} \cos \frac{\pi s}{2} \zeta(1+s) \Gamma(1-s)\Gamma(1+s) \phi(s) \\
     &=& \frac{i}{2} \int_{c-i\infty}^{c+i\infty} \frac{ds}{2 \pi}\, (4g)^{s-1} \frac{s \pi}{\sin \frac{\pi s}{2}} \zeta(1+s) \phi(s)  \label{fg}
\eeqa
The integral is independent of $c$ in the interval $0<c<1$. Now we need to consider $\phi(s)$. From the definition (\ref{phidef}) we find
\beq
|\phi(i\eta+c)| \le \int_0^{\pi} d\mu \int_0^{\pi} d\nu  \left(\sin\mu\sin\nu\right)^{c} |F(\mu,\nu)| = M, \ \ \ \forall \eta\in \mathbb{R}.
\label{phibound}
\eeq
where we assume that the integral on the right hand side is finite. For any given $F(\mu,\nu)$ this should be checked. It follows that
\beq
|f(g)| \le \frac{\pi M}{2} (4g)^{c-1} \int_{-\infty}^{+\infty} \frac{d\eta}{2 \pi}\,
    \sqrt{\frac{\eta^2+c^2}{\sinh^2 \frac{\pi \eta}{2}+\sin^2\frac{\pi c}{2}}} |\zeta(1+i\eta+c)|
\label{fbound}
\eeq
From the properties of the $\zeta$-function we know that $|\zeta(1+i\eta+c)| $ is bounded and goes to one for $\eta\rightarrow\pm\infty$.
We have then proven that, under the assumptions in eq.(\ref{phibound}), the integral defining $f(g)$ is absolutely convergent for any value of $g>0$.

This is quite generic, in our case, from eqs.(\ref{genfun}), (\ref{gen2}) and(\ref{cct}) we find that
$\tilde{c}_{m \bar{m}}$ can be expressed as
\begin{equation}
\tilde{c}_{m,\bar{m}}(g)= -4  g^2 \int_{0}^{\pi}\frac{d \mu}{\pi} \int_{0}^{\pi}\frac{d \nu}{\pi}
\frac{(\sin^2 \mu - \sin^2 \nu)}{(m-\bar{m})(m+\bar{m}+1)} e^{- (2 m+1)i \mu} e^{-(2 \bar{m}+1)i\nu} \mbox{Im}\psi'(1+ i \bar{g})
\end{equation}
with $\bar{g} = 4 g \sin\mu\sin\nu$. Comparing with (\ref{fdef}) we define
\begin{equation}
\tilde{F}_{m\bar{m}}(\mu,\nu)= -\frac{4g^2}{\pi^2}\frac{\sin^2 \mu -\sin^2 \nu}{\sin \mu \sin \nu} \frac{e^{-(2m+1) i \mu} e^{-(2\bar{m}+1) i \nu}}{(m-\bar{m})(m+\bar{m}+1)}
\end{equation}
and
\begin{equation}
\tilde{\phi}_{m\bar{m}}(s)=  \int_{0}^{\pi}  \int_{0}^{\pi} d \mu d \nu (\sin \mu \sin \nu)^s \tilde{F}_{m,\bar{m}}(\mu,\nu) \label{phi1}
\end{equation}
which is well defined for $\mbox{Re}(s)>0$. In fact we can evaluate it to be
\beq
\tilde{\phi}_{m\bar{m}}(s) = \frac{g^2 (-1)^{m+\bar{m}}}{4^{s-1} s (s+2)}
                    \frac{\Gamma(s+1)\Gamma(s+3)}
                        {\Gamma(\frac{s}{2}{\scriptstyle+1 -  m})\Gamma(\frac{s}{2}{ \scriptstyle + 2 + m})
                          \Gamma(\frac{s}{2}{\scriptstyle+1 -  \bar{m}})\Gamma(\frac{s}{2} {\scriptstyle + 2 + \bar{m}})}
\label{phitdef}
\eeq
For later use we notice the large $s$ behavior of $\tilde{\phi}_{m\bar{m}}(s)$ which is
\beq
\tilde{\phi}_{m\bar{m}}(s) =
\left\{\begin{array}{lcl} \frac{32g^2(-)^{m+\bar{m}}}{\pi s^3} &\mbox{\ \ for\ \ }& s\rightarrow\infty, \ \ (|\mbox{Arg}(s)|<\pi) \\ \\
          -\frac{32g^2(-)^{m+\bar{m}}}{\pi s^3} \tan^2\frac{\pi s}{2}&\mbox{for}& s\rightarrow\infty, \ \ (\mbox{Arg}(s)\neq0)    \end{array}    \right.
\label{phiap}
\eeq
The functions $\tilde{c}_{m,\bar{m}}$ can be written as $(0<c<1)$:
\beq
\tilde{c}_{m,\bar{m}}(g)=\half i g (-1)^{m+\bar{m}}  \int_{c-i\infty}^{c+i\infty} \frac{ds}{2 \pi}\, g^{s} \frac{(1+s) \pi}{\sin \frac{\pi s}{2}}
\zeta(1+s) \frac{\Gamma^2(1+s)}{\Gamma(\frac{s}{2}{\scriptstyle+1 -  m})\Gamma(\frac{s}{2}{ \scriptstyle + 2 + m})
                          \Gamma(\frac{s}{2}{\scriptstyle+1 -  \bar{m}})\Gamma(\frac{s}{2} {\scriptstyle + 2 + \bar{m}})}  \label{exp}
\eeq
which is the main result of our paper. As mentioned, it is an absolutely convergent integral for any real value of $g$, in fact, it converges exponentially at infinity
so it is a very good expression for numerical evaluation (for such purpose we can take \eg\ $c=\half$). Moreover, as seen below, it can be trivially expanded at large and small coupling by shifting the contour of integration
to the right and to the left. We derived (\ref{exp}) from the generating function (\ref{gen2}) but it also can be derived from (\ref{bessel}) by using a similar procedure\footnote{We thank G. Korchemsky for pointing this out to us based on his unpublished notes together with B.~Basso and J.~Kotanski.}. In fact after submitting this paper we learnt\footnote{We thank A. Kotikov for pointing out reference \cite{kl}.} that, in  \cite{kl}, it was already shown that an expression similar to (\ref{exp}) leads to the correct weak coupling expansion. However, in \cite{kl}, the strong coupling expansion was treated differently. It was considered as a formal series which was defined to be (\ref{exp}) and therefore an expansion such as (\ref{lec}) was not derived.  

\section{Small and large coupling expansions for $\tilde{c}_{m\bar{m}}$}

In this section we expand the expression (\ref{exp}) for the coefficients at large and small $g$. We start by proposing a method to expand expressions of the type (\ref{fg})
and apply it to a simple example in order to check and illustrate the idea. After that we apply it to the case of interest.

\subsection{General considerations} \label{sec:gc}
 To extract the weak coupling expansion ($g\rightarrow 0$) of $f(g)$ as given in (\ref{fg}) we shift the contour by an integer $K_1$ to the right, as shown in figure \ref{poles}. The calculation is similar to the one in Appendix A of \cite{kl}. 
From eq.(\ref{phibound}), and since $0<\sin\mu\sin\nu<1$ we find $|\phi(i\eta+c+K_1)|<M$. The same analysis that leads to the bound (\ref{fbound})
gives now that the integral over the lines extending from $c+i\eta$ to $K_1+i \eta$ with $\eta$ fixed (dashed line in the figure) go to zero exponentially
as $\eta\rightarrow\pm\infty$, so they can be ignored. Therefore we find that the shifted and original integral differ by (minus) the sum of the residues of
all the poles crossed, namely,
\beq
f(g) = \half \sum_{k} \mbox{Res}\left[(4g)^{s-1} \frac{s \pi}{\sin \frac{\pi s}{2}} \zeta(1+s) \phi(s), s=s_k\right] + R(K_1)\label{weak}
\eeq
where the remainder is given by the function $R(K)$ defined as
\beq
R(K) = \frac{i}{2} \int_{K+c-i\infty}^{K+c+i\infty}  \frac{ds}{2 \pi}\, (4g)^{s-1} \frac{s \pi}{\sin \frac{\pi s}{2}} \zeta(1+s) \phi(s)  \label{Rdef}
\eeq
where $K_1 \geq 1$ and the sum is over all poles of the integrand $s_k$ such that $c<\mbox{Re}(s_k)<c+K_1$. Typically, from the denominator
$\sin \frac{\pi s}{2}$ we will have poles for positive even $s$. This gives an expansion in odd powers of the coupling.

\begin{figure}
\epsfig{file=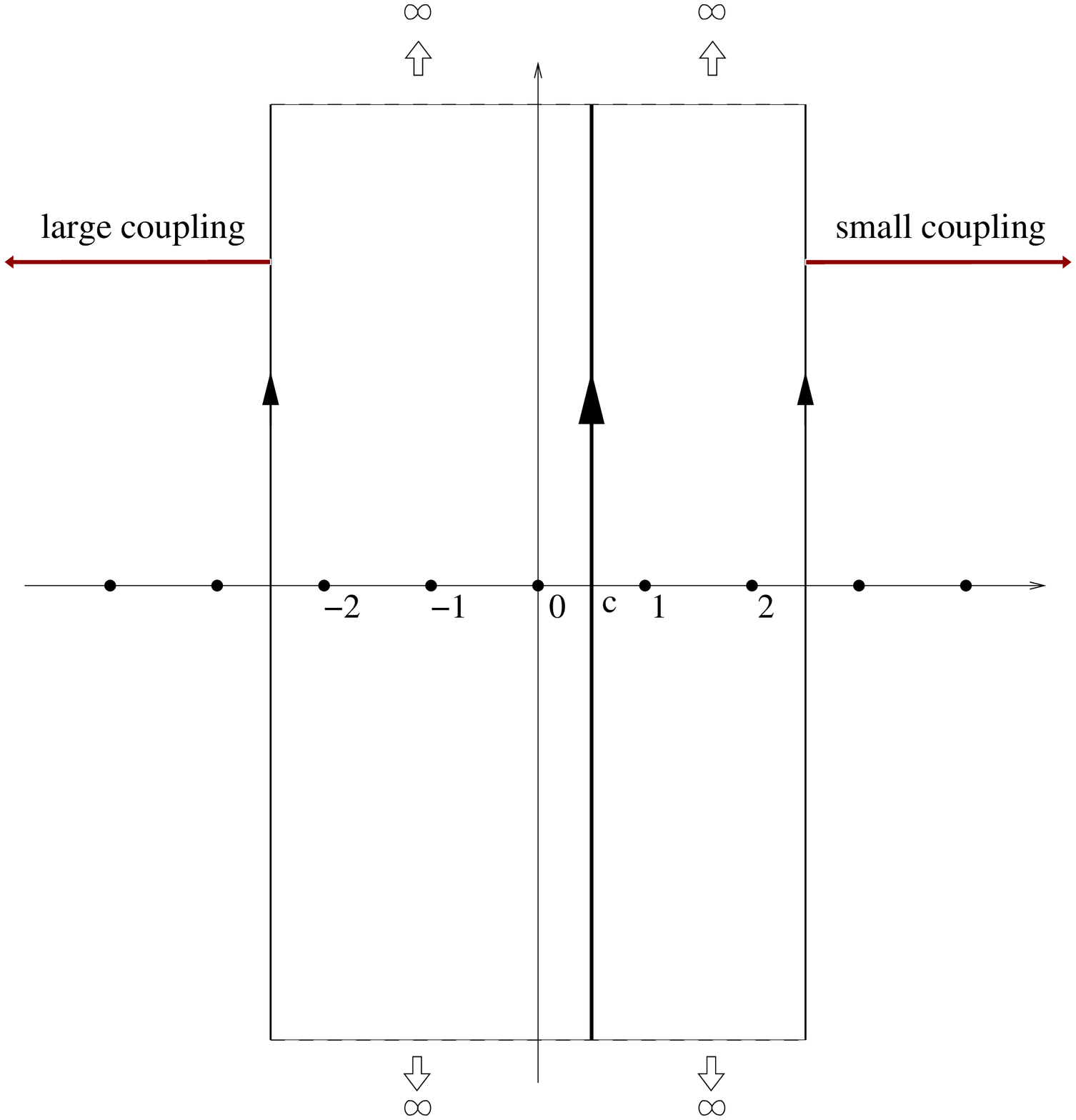, height=10cm}
\caption{Complex $s$-plane. The contour of integration is $(c-i\infty,c+i\infty)$. Shifting it to the right we obtain the weak coupling expansion
and shifting it to the left the strong coupling expansion.}
\label{poles}
\end{figure}

If the residual integral vanishes as $K_1\rightarrow\infty$ then we find an expression for $f(g)$ as an infinite series in powers of $g$. We are going
to see later that in the case of interest this only happens if $g<\frac{1}{4}$.

Now, more interestingly, we shift the contour by an integer $K_2>0$ to the left. To discard the integrals for large fixed imaginary part
we need to check later that the specific $\phi(s)$ that we have stays bounded in that region. Then those integrals also vanish exponentially
when taken to infinity. What we obtain is therefore
\beq
f(g) =  -\half \sum_{k} \mbox{Res}\left[(4g)^{s-1} \frac{s \pi}{\sin \frac{\pi s}{2}} \zeta(1+s) \phi(s), s=s_k\right] + R(-K_2)
 \label{strong}
\eeq
 where $s_k$ are the poles of the integrand such that $c-K_2<\mbox{Re}(s_k)<c$. The remainder $R(-K_2)$ is given by the same function defined in (\ref{Rdef}). Typically, for the $\phi(s)$ that we consider we have poles at negative
even $s$ from the denominator $\sin \frac{\pi s}{2}$ but also for negative odd $s$ from $\phi(s)$. This results in the strong coupling expansion
containing odd and even powers of $g$. Notice also the minus sign in front of the sum over poles since now the poles are crossed form right to left.
This method actually explains the conjecture in \cite{bes} that the coefficients of the small and large coupling expansions are related by an analytic
continuation and an extra minus sign. Although this is not always the case, for certain functions $\phi(s)$, the residues are simply related by
analytic continuation and the minus sign, as mentioned, is from closing the contour to the right or left.

  As we describe below, for the $\phi(s)$ of interest the integral on the right hand side in (\ref{strong}) does not vanish as $K_2\rightarrow\infty$.
In fact it gets smaller as $K_2$ increases up to a minimum point and then starts growing if we go further. Thus, we get an asymptotic series for
the expansion of $\tilde{c}_{m\bar{m}}(g)$. The integral is an exact expression for the residue of the series. This procedure is actually
similar in spirit to the integration by parts technique sometimes used to obtain asymptotic series.  To validate this approach now
we are going to study a simple example where we can check the result since we can find the same asymptotic series by using a different procedure.

\subsection{A simple example}

 A very simple example that allow us to illustrate the technique just described is provided by the function
\beq
\phi(s) = \frac{1}{(s+m)^2}
\eeq
where $m=2p+1$ is a positive odd integer. We then have
\beq
f(g) =  \frac{i}{2} \int_{c-i\infty}^{c+i\infty} \frac{ds}{2 \pi}\, (4g)^{s-1} \frac{s \pi}{\sin \frac{\pi s}{2}} \frac{\zeta(1+s)}{(s+m)^2}
\eeq
For positive real part of $s$ the only poles of the integrand in (\ref{fg}) are at even $s$. We find
\beqa
f(g) &=& \half \sum_{k=1}^\infty \mbox{Res}\left[(4g)^{s-1} \frac{s \pi}{\sin \frac{\pi s}{2}} \zeta(1+s) \phi(s), s=2k\right] \\
     &=& 2 \sum_{k=1}^\infty k (4g)^{2k-1} (-)^k \frac{\zeta(1+2k)}{(m+2k)^2}
\eeqa
For $\mbox{Re}(s)<\half$ the poles are at negative even $s$, at $s=0$ and at $s=-m$. Since $m$ is odd the pole at $s=-m$ is a simple pole since,
in that case, $\zeta(1-m)=0$. We find
\beqa
f(g) &=& -\frac{1}{4gm^2}+ 2 \sum_{k=1}^{K_2} k (4g)^{-2k-1} (-)^k \frac{\zeta(1-2k)}{(m-2k)^2} - \frac{m\pi}{2} (-)^{\frac{m-1}{2}} (4g)^{-m-1} \zeta'(1-m) \\
     &&     + \frac{i}{2} \int_{-K_2+c-i\infty}^{-K_2+c +i\infty} \frac{d s}{2 \pi}\, (4g)^{s-1} \frac{s\pi}{\sin \frac{s\pi}{2}}
            \frac{\zeta(1+s)}{(m+s)^2}
\label{sel}
\eeqa
We see a typical result. Odd powers appear but there is one term  proportional to $ g^{-m-1}$, namely an even power ($m$ is odd).
A more detailed analysis similar to what we present below  shows that we cannot take $K_2\rightarrow\infty$ since the integral grows for very large $K_2$.
For given $g\gg1$ an optimal value of $K_2$ exists such that the integral is at a minimum and can be discarded without significant error. Such error
however is perfectly quantified since one can put bounds to the residual integral or it can be evaluated numerically.
The point of this calculation is that in the appendix we compute this expansion using a different technique and obtaining the same result.
It is also useful to check numerically that the expansions are actually quite a good  approximation.

It is illustrative to consider also the case where $m$ is even. In that case there is a triple pole at $s=-m$ and we obtain a term proportional to $\ln g$ and $(\ln g)^2$. It can be worked out
but we leave it as an exercise for the interest reader.

\subsection{Small coupling expansion of $c_{m,\bar{m}}(g)$}

We can now apply the technique to the function $\tilde{\phi}_{m\bar{m}}$ defined in (\ref{phitdef}). For weak coupling we close the contour on the right hand side and pick up the simple poles
at $s=2 k$, with integer $k \geq 1$. The residues at those poles give the weak coupling expansion
\begin{equation}
\tilde{c}_{m,\bar{m}}(g)= (-1)^{m+\bar{m}} \sum_{k=1}^{\infty}g^{2 k+1} \frac{(2k+1) \zeta(2k+1) \cos(\pi k) \Gamma^2 (2 k+1)}{\Gamma(k+1-m)\Gamma(k+2+m)\Gamma(k+1-\bar{m})\Gamma(k+2+\bar{m})}  \label{qho}
\end{equation}
which agrees precisely with (\ref{CS}). Interestingly, the required extra minus sign just comes from closing the contour on the right.
The result in (\ref{qho}) is not the full result, we need to add to it the contribution from the second term in (\ref{weak}). Using the approximation
(\ref{phiap}), and assuming for simplicity that $K_1$ is even, we find for $R(K_1)$:
\beq
R(K_1) \simeq i(4g)^{K_1+2} (-)^{\half K_1+m+\bar{m}} \int_{c-i\infty}^{c+i\infty}\frac{ds}{2\pi}\, \frac{K_1+s}{\sin\frac{\pi s}{2}}\, \frac{(4g)^{s-1}}{s^3}, \ \ \ (K_1\rightarrow\infty)
\eeq
which clearly vanishes when $K_1\rightarrow\infty$ if $g<\frac{1}{4}$. Otherwise it diverges and consequently the series does not converge.

\subsection{Large coupling expansion of $c_{m,\bar{m}}(g)$}

As we already pointed out, to expand $c_{m, \bar{m}}$ at strong coupling we close the contour on the left side according to the discussion in
sec.\ref{sec:gc}. The poles are at $s=-n$ with $n=0,1,2,...$ and with a careful analysis of the residues we obtain precisely the coefficients expected
at strong coupling (\ref{CL}), which proves the conjecture in \cite{bes} for any $m,\bar{m}$. The analysis is slightly different for even and odd $n$.

The most straight-forward case is $s=-2 k-1$, $k\ge 1$, when there are simple poles due to the factor $\zeta(1+s) \Gamma^2 (1+s)$ which has
residue $\zeta'(-2 k)/((2 k)!)^2$. The result (\ref{CL}) is obtained directly, in fact for this case it is evident that the coefficients
are the analytical continuation of the small coupling ones.

At the particular value $s=-1$, the apparent double pole of $\zeta(1+s) \Gamma^2 (1+s)$ is reduced to a simple pole by the factor $(s+1)$.
Working out the residue it gives the coefficient
\begin{equation}
\bar{c}_{m, \bar{m}}^{(1)} = - \frac{1}{\pi}\frac{1}{(2 m+1) (2 \bar{m}+1)}
\end{equation}
which indeed is the right strong coupling coefficient (\ref{CL01}).

Now we consider the case $s=-2 k$, with $k \geq 0$ integer. Again we have only simple poles.
Let us first look at the pole at $s=0$. We observe that, in the denominator, $\Gamma(1- m)$ always has a pole for $m \geq 1$, which is the allowed
range for $m$. In addition if $\Gamma(1-\bar{m})$ also has a pole so there will be no overall pole at $s=0$ and the function is zero.
To have non-zero coefficients $\Gamma(1- \bar{m})$ has to have no pole. This sets $\bar{m}=0$. Computing the residue in this case
we obtain the coefficient proportional to $g$
\begin{equation}
c_{m,\bar{m}}^{(0)}= \delta_{\bar{m},0}
\end{equation}
which matches the corresponding strong coupling coefficient as obtained from string theory \cite{afs} (see also (\ref{CL01})).

For $s=-2 k$, $k=1,2,3,...$ we observe that both  $\Gamma(-k+1-m)$ and $\Gamma(-k+1 - \bar{m})$ always have poles for the range of interest
$m \geq 1$, $\bar{m}\geq 0$. The other two gamma functions in the denominator will not have poles for small enough values of $k$ such
that $-k +2 +\bar{m}>0$. Therefore, for $k< 2+ \bar{m}$ there is a simple pole at $s= - 2 k$. The residue is
\beq
 \mbox{Res}\bigg[\frac{\Gamma^2(1+s)}{\sin \frac{\pi s}{2} \ \Gamma(\frac{s+2 - 2 m}{2}) \ \Gamma(\frac{s+2 - 2 \bar{m}}{2})}, s=-2 k \bigg]
 = \frac{(-1)^{m+\bar{m}+k} \ \Gamma(m+k) \ \Gamma(\bar{m}+k)}{2 \pi \ \Gamma^2 (2 k)}
\eeq
Using this residue we find that the coefficients agree with the strong coupling expansion coefficients $c_{m,\bar{m}}^{(n)}$ obtained in \cite{bhl}
(see eq.(\ref{CL})). For $-k +2 +\bar{m} \leq 0$, the pole of $\Gamma(-k +2 +\bar{m})$ starts contributing resulting in no net pole.
Thus, the series terminates at $k=2+\bar{m}$, which again is in agreement with the strong coupling coefficients.

As in the case of weak coupling there is a remainder $R(-K_2)$. That integral gets smaller as we increase $K_2$ but after certain value of $K_2$ it
starts increasing regardless of what large the value of $g$. A lengthy but simple computation shows that such value is $K_2=8\pi g$ and therefore
the maximum precision that we can obtain at strong coupling is
\beq
R(-8\pi g) = (-)^{m+\bar{m}+1} \frac{5}{32\pi^3} \frac{e^{-8\pi g}}{g^{\frac{3}{2}}} \left(1+\mathcal{O}\left(\frac{1}{g}\right)\right)  \label{mp}
\eeq
where we only kept the leading term in the large $g$ expansion of $R(-8\pi g)$.

\bigskip

As we explain at the beginning of this section, this derivation proves the conjecture made in \cite{bes} that the coefficients of the strong and weak
coupling are simply related. This comes out naturally from the integral representation (\ref{exp}) of $c_{m, \bar{m}}$ by closing up the
contour differently for strong and weak coupling.

\section{Summing the dressing phase}

In this section we assume $|x_1|>1$, $|x_2| > 1$ as appropriate for the problem we are studying \cite{bes}. $x_1, x_2$ are $g$-dependent, however, here
we assume them fixed and consider only the $g$-dependence of the phase through $c_{m, \bar{m}}(g)$. One needs to consider further the $g$-dependence of $x_1, x_2$, which depending on
solution one is interested in may give different terms, such as $\ln g$ terms\footnote{We thank A. Tseytlin for pointing this out to us.} \cite{rs}.
Given that
\beqa
\chi(x_1,x_2,g) =-2 \sum_{\bar{m}=0}^{\infty}\sum_{m=\bar{m}+1}^{\infty} \frac{\tilde{c}_{m,\bar{m}}(g) }{x_1^{m-\bar{m}} x_2^{m+\bar{m}+1}}
\eeqa
and using eq.(\ref{cmmbar}) we obtain, by performing the sums explicitly\footnote{Although we followed a different route, we arrive at an expression very similar to the one obtained by Dorey, Hofman and Maldacena \cite{dhm}. In fact they are related by a change of variables. Another related integral representation was found in \cite{ksv}. The weak coupling expansion of the DHM representation was discussed in \cite{Bajnok}.
Our main contribution in this section is in how to systematically expand the function $\chi(x_1,x_2,g)$ at weak and strong coupling.}:
\begin{equation}
\chi(x_1,x_2)
= -2 x_2 \int_0^{\pi} \frac{d \mu}{\pi}\int_0^{\pi} \frac{d \nu}{\pi} \frac{e^{i(\mu+\nu)}}{(x_1 x_2 e^{2 i \mu}-1)(x_2^2 e^{2 i (\mu+\nu)}-1)}[\gamma \bar{g}+
\arg \Gamma(1+ i \bar{g})]  \label{qoi}
\end{equation}
It is more convenient to analyze the derivative
\begin{equation}
\partial^2_g \chi(x_1,x_2)=32  x_2 \int_0^{\pi} \frac{d \mu}{\pi}\int_0^{\pi} \frac{d \nu}{\pi} \frac{e^{ i (\mu+\nu)}(\sin \mu \sin \nu)^2}{(x_1 x_2 e^{2 i \mu}-1)(x_2^2 e^{2 i (\mu+\nu)}-1)}\mbox{Im}[\psi'(1+ i \bar{g})]
\end{equation}
which has the form (\ref{fdef}) with
\beq
F_{\chi}(\mu,\nu) = \frac{32  x_2}{\pi^2}  \frac{e^{ i (\mu+\nu)} \sin \mu \sin \nu }{(x_1 x_2 e^{2 i \mu}-1)(x_2^2 e^{2 i (\mu+\nu)}-1)}
\label{Fchidef}
\eeq
We can now use the method described in the previous section to make a straight-forward small and large coupling expansion  of $\chi(x_1,x_2,g)$.

\subsection{Small coupling expansion}

 First we find the weak coupling expansion of $\chi$. This can be obtained by using the method of shifting the contour or by
simply expanding (\ref{qoi}) in small $g$
\begin{equation}
\chi(x_1,x_2) = \sum_{k=1}^{\infty}\chi^{(k)}(x_1,x_2) g^{2 k+1}
\end{equation}
where
\begin{equation}
\chi^{(k)}(x_1,x_2) =\half \frac{(-1)^{k+1} 4^{2 k-1}  \zeta(2k+1) }{2 k+1}\, \phi(k)
\end{equation}
and
\begin{equation}
\phi(k)=32 x_2  \int_0^{\pi} \frac{d \mu}{\pi}\int_0^{\pi} \frac{d \nu}{\pi}\frac{(\sin \mu \sin \nu)^{2 k+1}e^{i (\mu+\nu)}}{(x_1 x_2 e^{2 i \mu}-1)(x_2^2 e^{2 i (\mu+\nu)}-1)}
\end{equation}
Using $y= e^{2 i \mu}, z=e^{2 i \nu}$ we can write this integral as a double integral over two unit circles
\begin{equation}
\phi(k)= \frac{2 x_2}{ \pi^2 4^{n}}\oint dy dz \frac{(y-1)^{n+1} (z-1)^{n+1}}{y^{\frac{n}{2}+1} z^{\frac{n}{2}+1}}\frac{1}{(x_1 x_2 y -1)(x_2^2 y z - 1)}
\label{chismall}
\end{equation}
where $n=2k$ is the position of the pole.  We use $n$ instead of $k$ for later use, now we can replace $n=2k$.
The simplest way to evaluate this integral is to expand it over the domain outside the unit circles. The only poles are those at infinity.
For convenience we consider further $y \rightarrow 1/y$, and $z \rightarrow 1/z$, then the integral becomes
\begin{equation}
\phi(k)= \frac{2 x_2}{\pi^2 4^{2 k}}\oint dy dz \frac{(1-y)^{2 k+1} (1-z)^{2k+1}}{y^k z^{k+1}}\frac{1}{(x_1 x_2 -y)(x_2^2 - y z)}
\end{equation}
 Computing the residues of the poles at zero we obtain
\begin{equation}
\chi^{(k)}(x_1,x_2)=  \frac{2x_2 (-1)^{k+1} \zeta(2 k+1)}{2k+1}\frac{1}{k! (k-1)!}\frac{d^{k}}{d z^k} \frac{d^{k-1}}{d y^{k-1}}[\frac{(1-y)^{2 k+1} (1-z)^{2 k+1}}{(x_1 x_2 -y)(x_2^2- y z)}]\bigg|_{z=0,y=0}  \label{qko}
\end{equation}
The expansion coefficients at any order can be extracted right away from (\ref{qko}). For example the first ones are
\begin{equation}
\chi^{(1)}(x_1,x_2)=-2\frac{\zeta(3)}{x_1 x_2^2}, \quad \quad \chi^{(2)}(x_1,x_2)=2\frac{x_1-2 x_2 +10 x_1 x_2^2}{x_1^2 x_2^4}\zeta(5)
\end{equation}
We can write the result (\ref{qko}) in terms of a finite double sum as
\begin{equation}
\chi^{(k)}(x_1,x_2)= 2 \frac{ (-1)^{k+1} \zeta(2 k+1)}{(2k+1)x_2^{2 k+1}}\sum_{p=0}^{q-1} \sum_{q=0}^k \bino{2k+1}{p}\bino{2k+1}{q} (-x_1 x_2)^p \left(-\frac{x_2}{x_1}\right)^q
\end{equation}

\subsection{Large coupling expansion, odd coefficients}

To perform the strong coupling expansion let us use the same method as in the previous section. Analyzing eq.(\ref{strong}) for $s=-2k-1$,
$k=0,1,2, \ldots$ the factor in front of $\phi(s)$ actually vanishes because $\zeta(-2k)=0$. It turns out, however, that $\phi(s)$ has double poles
there. This is generic and can be seen by rewriting eq.(\ref{phidef}) as
\beq
\phi(s) = \int_0^\pi d\mu \int_0^\pi d\nu \,\mu^s \nu^s \left(\frac{\sin\mu\sin\nu}{\mu\nu}\right)^s F(\mu,\nu)
\eeq
When $s$ gets close to a negative integer $-n$ we can expand
\beq
\left(\frac{\mu\nu}{\sin\mu\sin\nu}\right)^n F(\mu,\nu) = \sum_{l,l'} F_{l,l'} \mu^{l} \nu^{l'}
\eeq
and perform the integrations
\beq
\phi(s) = \sum_{l,l'}  F_{l,l'} \frac{\pi^{l+l'+2s+1}}{(l+s+1)(l'+s+1)}
\eeq
which can now be extended to arbitrary values of $s$. It is clear that there are double poles at negative integers $s=-n$ and the coefficient is
simply $F_{(n-1),(n-1)}$. Therefore, the coefficients of the double poles are the diagonal coefficients of the double Taylor expansion.
 There is one point still to take into account which is that there are potential contributions also from the $\mu=\pi$ or $\nu=\pi$ limits. In fact
it is convenient to express everything in terms of integrals from $0$ to $\frac{\pi}{2}$ only
\begin{equation}
\partial^2_g \chi(x_1,x_2)= \int_0^{\frac{\pi}{2}} d \mu \int_0^{\frac{\pi}{2}} d \nu \,
 \sin \mu \sin \nu\, \mbox{Im}[\psi'(1+ i \bar{g})]B_{\chi}(\mu,\nu)
\end{equation}
where
\beq
B_{\chi}(\mu,\nu) =F_{\chi}(\mu,\nu)+F_{\chi}(-\mu,\nu)+F_{\chi}(\mu,-\nu)+F_{\chi}(-\mu,-\nu)
\eeq
with $F_{\chi}$ as defined in (\ref{Fchidef}). When expanding in powers of $\mu,\nu$ we only get even powers which then means that the
double poles are at odd values of $n$. This explains why we only get even powers of the coupling this way. For odd powers (namely $s$ even),
there are poles in the factor in front of $\phi(s)$, not in $\phi(s)$ itself. Going back to the odd powers we can simply write for the coefficient
\beq
\partial_g^2\chi^{(-2k-1)}(x_1,x_2) = -2 (4g)^{-2k-2} (2k+1) \pi (-)^k \zeta'(-2k) \bar{\phi}(-2k-1)
\eeq
where $\bar{\phi}(-2k-1)$ is the coefficient of the double pole which as we just said can be computed by a double Taylor expansion. There is a factor of
four in front since each term in $B_{\chi}(\mu,\nu)$ contributes the same.
More precisely, extending $\mu$, $\nu$ to complex values we can compute it as
\beq
\bar{\phi}(-2k-1) =  \frac{128 x_2}{\pi^2}\frac{1}{(2\pi i)^2} \oint_{\mathcal{C}_0} \frac{d\mu}{\mu^{2k+1}} \oint_{\mathcal{C}_0} \frac{d\nu}{\nu^{2k+1}}
                     \left(\frac{\mu\nu}{\sin\mu\sin\nu}\right)^{2k+1}
                        \frac{e^{ i (\mu+\nu)} \sin \mu \sin \nu }{(x_1 x_2 e^{2 i \mu}-1)(x_2^2 e^{2 i (\mu+\nu)}-1)}
\eeq
where ${\mathcal{C}_0}$ denotes a small contour surrounding the origin. It is convenient to change variables to $y=e^{2i\mu}$, $z=e^{2i\nu}$ such that
\beq
\bar{\phi}(-2k-1) =
-4 \frac{2 x_2}{ \pi^2 4^{n}}\frac{1}{(2\pi i)^2}\oint_{\mathcal{C}_1} dy dz \frac{(y-1)^{n+1} (z-1)^{n+1}}{y^{\frac{n}{2}+1}
             z^{\frac{n}{2}+1}}\frac{1}{(x_1 x_2 y -1)(x_2^2 y z - 1)}
\label{phiodd}
\eeq
where $n=-2k-1$. We write it in this way to show that the integrand is similar to the one at small coupling (\ref{chismall}).
The difference is that the integral is done now over contours surrounding $y=z=1$ whereas before they where at infinity. To obtain an explicit expression
is now convenient to change variables to $\omega_1=y-1$, $\omega_1=(z-1)y$ and then expand the integrals in powers of $\omega_{1,2}$ to get
\beq
\bar{\phi}(-n) = -\frac{4^{n+2}}{2\pi^2 x_1 x_2^2} \sum_{a=0}^{n-2} \sum_{b=0}^a\sum_{c=0}^{2n-4-a}
               \bino{n-2}{a-b}\bino{\half n-1}{2n-4-a-c}\bino{2n-4-a}{n-2-a}
               \left(\frac{x_1x_2}{1-x_1x_2}\right)^{b+1}\left(\frac{x_2^2}{1-x_2^2}\right)^{c+1}
\eeq
where we now used $n=2k+1$. Notice that all sums are over a finite range. The final formula is therefore
\beqa
\lefteqn{\chi^{(-n)}(x_1,x_2) = \frac{1}{\pi} g^{-n-1}  (-)^{\frac{n-1}{2}} \frac{\zeta'(1-n)}{n-1}  \frac{1}{x_1 x_2^2} }\ \ \ \ \  &&  \\
                     && \sum_{a=0}^{n-2} \sum_{b=0}^a\sum_{c=0}^{2n-4-a}
               \bino{n-2}{a-b}\bino{\half n-1}{2n-4-a-c}\bino{2n-4-a}{n-2-a}
               \left(\frac{x_1x_2}{1-x_1x_2}\right)^{b+1}\left(\frac{x_2^2}{1-x_2^2}\right)^{c+1}  \label{kio}
\eeqa
for $n$ odd. In the last step we integrated twice over $g$ since we were computing $\partial_g^2 \chi(x_1,x_2,g)$. The expression (\ref{kio}) is not valid for $n=1$; however, this case can be considered separately and summations can be performed \cite{af}. Although we derived this last expression for $n$ odd,
it can be seen to also capture the correct $x_{1,2}$ dependence for $n$ even as we discuss below.

\subsection{Large coupling expansion, even coefficients}

The results in the previous section can be obtained in a simpler way for even $n=2k$. We take $k\geq 1$; the case $n=0$ can be treated separately and the result is well known \cite{af}. Using (\ref{CL}) we can write
\begin{equation}
\chi^{(-2k)}(x_1,x_2)=-\frac{\zeta(2k)}{x_2 (-2 \pi)^{2 k} \Gamma(2k-1)}\sum_{\bar{m}=0}^{\infty}\sum_{m=\bar{m}+1}^{\infty}\frac{\Gamma(m+k)}{\Gamma(m+2-k)}\frac{\Gamma(\bar{m}+k)}{\Gamma(\bar{m}+2-k)}
\frac{1}{(x_1 x_2)^m}\left(\frac{x_1}{x_2}\right)^{\bar{m}}
\end{equation}
As in section 3 where we summed the weak coupling expansion, here we can extend the sum and compute
\begin{equation}
\tilde{D}^{(-2 k)}(y,z) \equiv \sum_{\bar{m}=0}^{\infty}\sum_{m=0}^{\infty}\tilde{c}^{(-2k)}_{m, \bar{m}} y^m z^{\bar{m}}=
\frac{\zeta(2k) \Gamma(2k-1)}{2 (-2 \pi)^{2 k}}(y-1)^{1-2k} y^{k-1} (z-1)^{1-2 k} z^{k-1}
\end{equation}
In order to perform the sums above we took $|y|<1$, $|z|<1$. We can then obtain back the coefficients $\tilde{c}^{(-2k)}_{m, \bar{m}}$ from
\begin{equation}
\tilde{c}^{(-2k)}_{m, \bar{m}}=-\frac{1}{4 \pi^2}\oint d y d z y^{-m-1} z^{-\bar{m}-1}\tilde{D}^{(-2 k)}(y,z)
\end{equation}
where the integrals are over circles with radius smaller than unity. Replacing in  $\chi^{(-2k)}(x_1,x_2)$ and performing the sums over $m, \bar{m}$ we obtain
\begin{equation}
\chi^{(-2k)}(x_1,x_2)= \frac{x_2 \zeta(2k) \Gamma(2k-1)}{4 \pi^2 (-2 \pi)^{2 k}}\oint d y dz \frac{(y-1)^{1-2k} y^{k-1}(z-1)^{1-2k} z^{k-1}}{(x_1 x_2 y-1)(x_2^2 y z -1)}
\end{equation}
Comparing with eq.(\ref{phiodd}) and since there are no poles at infinity, we see why the result (\ref{kio}) has the right $x_{1,2}$ dependence for $n$
even. In this case, however\footnote{For $n$ odd there are cuts so the following procedure does not give a simpler answer.}, we can find a simpler expression if we perform first the integral in $z$. We only have a simple pole at $z= \frac{1}{x_2^2 y}$. Computing the residue at the pole we obtain an integral over $y$
\begin{equation}
\chi^{(-2k)}(x_1,x_2)= \frac{i x_2^{2k-1} \zeta(2k) \Gamma(2k-1)}{2 \pi (-2 \pi)^{2k}}\oint dy \frac{y^{2k-2} (y-1)^{1-2k}}{(x_1 x_2 y-1)(1-x_2^2 y)^{2k-1}}
\end{equation}
There is a simple pole at $y=\frac{1}{x_1 x_2}$ and a pole of order $2k-1$ at $y=\frac{1}{x_2^2}$. Computing the residues we obtain
\begin{eqnarray}
 \chi^{(-2k)}(x_1,x_2)&=& - \frac{\zeta(2k) \Gamma(2k-1)x_2^{2k-1}}{(-2 \pi)^{2k}}\bigg[(1-x_1 x_2)^{1-2k} (1-\frac{x_2}{x_1})^{1-2k}\nonumber\\
 &+&
 \frac{1}{x_2^{4k-2}(2k-2)!}\frac{d^{2k-2}}{d y^{2k-2}} \frac{y^{2k-2} (y-1)^{1-2k}}{1- x_1 x_2 y} \bigg|_{y=\frac{1}{x_2^2}}\bigg]
\end{eqnarray}
We can further express the derivative term in terms of a finite double sum
\begin{eqnarray}
\chi^{(-2k)}(x_1,x_2)&=& - \frac{\zeta(2 k) \Gamma(2k-1) x_2^{2k-1}}{(-2 \pi)^{2 k}}\bigg[(1- x_1 x_2)^{1-2 k} (1-\frac{x_2}{x_1})^{1-2 k}\\
&-& \frac{x_2 (x_2^2-1)^{1-2 k}}{x_2-x_1}\sum_{p=0}^{2k-2}\sum_{m=0}^p  \bino{2k-2}{p}\bino{2k-2+m}{m} \left(\frac{x_1}{x_2-x_1}\right)^p
\left(\frac{x_1 (x_2^2-1)}{x_2-x_1}\right)^{-m}\bigg] \nonumber
\end{eqnarray}
The result can be extracted right away for any $k$. For example $\chi^{(-2)}(x_1,x_2)$ is
\begin{equation}
\chi^{(-2)}(x_1,x_2)=-\frac{x_2}{24 (x_1 x_2-1)(x_2^2-1)}
\end{equation}
The results for $ \chi^{(-2k)}(x_1,x_2)$ that we obtain match the corresponding expressions obtained in \cite{bhl}.

\subsection{Large coupling expansion, exponential terms}

For the maximum precision at strong coupling we found in (\ref{mp}) that $N=8 \pi g$. Thus $N=8 \pi g$ terms are to be summed in this case. The remainder $R_N$ can be summed over $m, \bar{m}$ and we obtain exponential corrections
\begin{equation}
R_N= - \frac{5}{16 \pi^3} \frac{x_2}{(1+x_1 x_2)(x_2^2-1)}\frac{e^{-8 \pi g}}{g^{\frac{3}{2}}} \left(1+\mathcal{O}\left(\frac{1}{g}\right)\right)
\end{equation}
Such exponential corrections appear in the all loop Bethe Ansatz, therefore they need to be included in the study of any particular solution.

\section{Summary and Outlook}

 In this paper we have analyzed the function $\chi(x_1,x_2,g)$ that enters in the all-loop Bethe Ansatz for the $SL(2)$ sector of
$\mathcal{N}=4$ SYM theory. We found that the coefficients of its expansion in inverse powers of $x_1$ and $x_2$ can be written as
\beq
\tilde{c}_{m,\bar{m}}(g)=\half i g (-1)^{m+\bar{m}}  \int_{c-i\infty}^{c+i\infty} \frac{ds}{2 \pi}\, g^{s} \frac{(1+s) \pi}{\sin \frac{\pi s}{2}}
\zeta(1+s) \frac{\Gamma^2(1+s)}{\Gamma(\frac{s}{2}{\scriptstyle+1 -  m})\Gamma(\frac{s}{2}{ \scriptstyle + 2 + m})
                          \Gamma(\frac{s}{2}{\scriptstyle+1 -  \bar{m}})\Gamma(\frac{s}{2} {\scriptstyle + 2 + \bar{m}})}
\eeq
where $0<c<1$, which are well defined expressions for any $g$ real and positive. Moreover, by shifting the contour to the right or to the left,
we obtain an expansion for small or large $g$ that agrees with the known expressions. This proves the conjectured equivalence of both 
expansions. For clarity we remind the reader that these coefficients are usually denoted as $c_{r,s}=2(r-1)(s-1) \tilde{c}_{r,s}$ and $r,s$ are related to
$m$, $\bar{m}$ by $s=m+\bar{m}+2$, $r=m-\bar{m}+1$. The large coupling expansion is only asymptotic and we find that it is convenient to sum a
number $K=8\pi g$ of terms. In that case, the remainder, that we compute explicitly, can be estimated to be
$R_K \simeq (-)^{m+\bar{m}+1} \frac{5}{32\pi^3} \frac{e^{-8\pi g}}{g^{\frac{3}{2}}}$. The computation of the remainder gives a precise meaning to the
strong coupling expansion whose status was unclear in previous work \cite{kl,fr}.

We then considered the expansion of $\chi(x_1,x_2,g)$ in powers of $g$ for large and small coupling. We obtain that, for small coupling
\beq
\chi(x_1,x_2,g) = \sum_{k=1}^\infty \half \frac{(-)^{k+1} 4^{2k-1}\zeta(2k+1)}{2k+1} \phi(k) g^{2k+1}
\eeq
where
\beq
\phi(k)= \frac{2 x_2}{ \pi^2 4^{n}}\oint_{\mathcal{C}_\infty} dy dz \frac{(y-1)^{n+1} (z-1)^{n+1}}{y^{\frac{n}{2}+1} z^{\frac{n}{2}+1}}\frac{1}{(x_1 x_2 y -1)(x_2^2 y z - 1)}
\eeq
with $n=2k$ and the integrals are done for a contour around infinity.
For large coupling we get an asymptotic expansion
\beq
\chi(x_1,x_2,g) = \sum_{n=0}^N \chi^{(-n)}(x_1,x_2) g^{1-n} + R_N
\eeq
where, for odd $n$ we find
\beq
\chi^{-2k-1}(x_1,x_2) = -2\ 4^{-2k-2} (2k+1) \pi (-)^k \zeta'(-2k) \phi(-2k-1)
\eeq
where
\beq
\phi(n) =-4 \frac{2 x_2}{ \pi^2 4^{n}}\frac{1}{(2\pi i)^2}\oint_{\mathcal{C}_1} dy dz \frac{(y-1)^{n+1} (z-1)^{n+1}}{y^{\frac{n}{2}+1}
             z^{\frac{n}{2}+1}}\frac{1}{(x_1 x_2 y -1)(x_2^2 y z - 1)}
\eeq
which is similar to the small coupling expression but with the contours now surrounding the point $y=z=1$. Finally, for even $n$ we find
\beq
\chi^{(-n)}(x_1,x_2)= \frac{x_2 \zeta(n) \Gamma(n-1)}{4 \pi^2 (-2 \pi)^{n}}\oint_{\mathcal{C}_{1^-}} d y dz \frac{(y-1)^{1-n} y^{\frac{n}{2}-1}(z-1)^{1-n} z^{\frac{n}{2}-1}}{(x_1 x_2 y-1)(x_2^2 y z -1)}
\eeq
where $n=2k$ and the contours are circles around the origin with radius slightly smaller than $1$. All the residues can be evaluated very
simply in terms of finite sums of combinatorial symbols and appropriate functions of $x_1$, $x_2$.  Many of these functions were known but,
to our knowledge, no generic expression was given for all of them such as the one we present here. We hope that the use of these results
would simplify the study of the all-loop Bethe Ansatz and its small and large coupling expansions.

\section{Acknowledgments}

We are  grateful to G. Korchemsky, A. Kotikov and  A. Tseytlin  for useful comments on the first version of this paper. 
This work was supported in part by NSF under grant PHY-0805948, DOE under grant DE-FG02-91ER40681 and by the Alfred P. Sloan foundation.
M.K. is grateful to the Aspen Center for Physics for hospitality during the initial stages of this work.

\appendix
\subsection*{Appendix A:  A simple example of small and large coupling expansions.}

Suppose we have the function
\beq
f_m(g)  = - \int_0^1 du\, u^{m} \ln u \mbox{Im}\left[ \psi'(1+4igu) \right]
\eeq
and we want to expand it for small and large $g$. To use the method discussed in section \ref{sec:gc} we need to define
\beq
\phi(s) =  - \int_0^1 du\, u^s \ln u u^{m-1} =  \frac{1}{(m+s)^2}
\eeq
namely, precisely the function we used in our simple example at the end of that section.
Using that
\beq
\psi'(1+4igu) = \left. \int_0^\infty \frac{te^{-xt}}{1-e^{-t}}dt \right|_{x=1+4igu} = \int_0^\infty \frac{te^{-4igut}}{e^t-1}dt
\eeq
we can rewrite
\beq
f_m(g) = \int_0^\infty \sin(4g\zeta) H(\zeta)
\eeq
where we changed variables to $t\rightarrow\zeta=ut$ and defined
\beq
H(\zeta) = \zeta\int_0^1 \frac{du\,u^{m-2}\ln u}{e^{\frac{\zeta}{u}}-1} = -\zeta\int_1^\infty \frac{dv}{v^{m}}\frac{\ln v}{e^{\zeta v}-1}
\eeq
 Therefore $f_m(g)$ is just the sine Fourier transform of $H$ and we want to expand it for large $g$. The result is determined by the behavior of $H(\zeta)$
around $\zeta=0$. To expand $H(\zeta)$ we need to first take a detour and remind ourselves some properties of the $\zeta$ function.
We have, for $Re(s)>1$
\beqa
\zeta(s) &=& \sum_{k=1}^\infty \frac{1}{k^s} = \sum_{k=1}^\infty \frac{1}{\Gamma(s)} \int_0^\infty \lambda^{s-1} e^{-k\lambda} d\lambda
                                             = \frac{1}{\Gamma(s)} \int_0^\infty \frac{\lambda^{s-1}\,d\lambda}{e^\lambda-1} \\
         &=& \frac{1}{\Gamma(s)} \int_0^a \frac{\lambda^{s-1}\,d\lambda}{e^\lambda-1}
            +\frac{1}{\Gamma(s)} \int_a^\infty \frac{\lambda^{s-1}\,d\lambda}{e^\lambda-1}
\eeqa
If $0<a<2\pi$ we can expand the integrand of the first integral by using the definition of the Bernoulli numbers $B_p$ to obtain
\beq
\zeta(s) = \frac{1}{\Gamma(s)} \sum_{p=0}^\infty \frac{B_p}{p!} \frac{a^{p+s-1}}{p+s-1}
            + \frac{a^s}{\Gamma(s)} \int_1^\infty \frac{\lambda^{s-1}\,d\lambda}{e^{a\lambda}-1}
\eeq
where we also changed variables $\lambda\rightarrow\frac{\lambda}{a}$ in the second integral. Although we derive this result in the region $Re(s)>1$, the right hand side defines
a meromorphic function of $s$ for any complex $s$. Since $\Gamma(s)$ never vanishes the last term is analytic. The sum in the first term has poles at $s=1,0$ and negative integers.
The poles go away since $\frac{1}{\Gamma(s)}$ has zeros at those points except $s=1$. We conclude that the right hand side has a single pole at $s=1$ with residue 1, a well-known property
of the $\zeta$-function. This is a useful way to find the analytic continuation and, in fact, we use a similar method in the main text.
The purpose of this exercise is that we derived the identity
\beq
\int_1^\infty \frac{\lambda^{s-1}\,d\lambda}{e^{a\lambda}-1} = a^{-s} \Gamma(s) \zeta(s) -  \sum_{p=0}^\infty \frac{B_p}{p!} \frac{a^{p-1}}{s+p-1}
\eeq
 Now we can expand around $s=-m+1+\epsilon$. If we identify the first order term in both sides we get:
\beqa
\lefteqn{\int_1^\infty \frac{\lambda^{-m}\ln\lambda\,d\lambda}{e^{a\lambda}-1} =  \sum_{\begin{array}{c}\scriptstyle p=0\\\scriptstyle  p\neq m\end{array}}^\infty \frac{B_p}{p!} \frac{a^{p-1}}{(p-m)^2}
-\frac{(-)^m a^{m-1}}{(m-1)!} \left\{ \half\zeta(1-m)(\ln a)^2 + \zeta'(1-m)\psi(m) \right.} \ \ \ \ \\
&& \left.  - \ln a\left[\zeta'(1-m)+\psi(m)\zeta(1-m)\right] +\half\zeta''(1-m) +\zeta(1-m)\left[\frac{\pi^2}{6}+\half\psi^2(m)-\half\psi'(m)\right]\right\} \nonumber
\eeqa
which gives the desired expansion for $H(\zeta)$. We are considering the case $m$ in which case $\zeta(1-m)=0$ and the expansion simplifies considerably:
\beq
H(\zeta) = - \sum_{\begin{array}{c}\scriptstyle p=0\\\scriptstyle  p\neq m\end{array}}^\infty \frac{B_p}{p!} \frac{\zeta^p}{(p-m)^2}
 -\frac{\zeta^m}{(m-1)!} \left[\half\zeta''(1-m)-\zeta'(1-m)\ln \zeta+\zeta'(1-m)\psi(m)\right]
\label{Hexp}
\eeq
To compute the Fourier transform we include a decreasing exponential which has no effect in the region $\zeta\rightarrow 0$ that we consider and get
\beqa
 \int_0^{\infty} \zeta^{s-1} \sin(4g\zeta) e^{-\epsilon\zeta} d\zeta &=& (4g)^{-s} \sin \frac{\pi s}{2}\Gamma(s) \ \ \ \ \ (\epsilon\rightarrow 0) \\
 \int_0^{\infty} \zeta^{s-1}\ln\zeta \sin(4g\zeta) e^{-\epsilon\zeta} d\zeta &=&
                (4g)^{-s}  \sin \frac{\pi s}{2}\Gamma(s) \left[-\ln(4g) + \psi(s)+\frac{\pi}{2} \mbox{cotan} \frac{\pi s}{2}\right] \ \ \ \ \ (\epsilon\rightarrow 0) \nonumber
\label{sineFeps}
\eeqa
Replacing in (\ref{Hexp}) we get
\beq
f_m(g) = - \sum_{p=0}^\infty B_{2p} (-)^p \frac{(4g)^{-2p-1}}{(2p-m)^2} -\frac{m\pi}{2}\zeta'(1-m) (4g)^{-m-1} \sin\frac{m\pi}{2}
\eeq
which in view of the identity $\zeta(-n)=-\frac{B_{n+1}}{n+1}$ , $n>0$, agrees with eq.(\ref{sel}) including the term $p=0$.

\end{document}